\documentclass[twocolumn,showpacs,preprintnumbers,amsmath,amssymb,floatfix]{revtex4}
\usepackage{graphicx}
\usepackage{dcolumn}

\newcommand{\be}{\begin{equation}}
\newcommand{\ee}{\end{equation}}
\newcommand{\ba}{\begin{eqnarray}}
\newcommand{\ea}{\end{eqnarray}}
\newcommand{\n}[1]{\label{#1}}
\newcommand{\eq}[1]{(\ref{#1})}
\newcommand{\hhh}{\, ,\hspace{0.2cm}}

\begin{document}
 
\title{Gregory-Laflamme instability of 5D electrically
charged black strings}
\author{Valeri P. Frolov}
\email{frolov@phys.ualberta.ca}
\author{Andrey A. Shoom}
\email{ashoom@phys.ualberta.ca}
\affiliation{Theoretical Physics Institute, University of Alberta, 
Edmonton, AB, Canada,  T6G 2G7}
\date{\today}

\begin{abstract} We study the Gregory-Laflamme instability of 5D compactified, electrically charged black strings. We consider static, linear perturbations of these solutions and derive master equation for these perturbations. Using numerical analysis we search for the threshold unstable modes and derive the critical wave-length. The results are illustrated in the black string mass v.s. electric charge phase diagram. Similar diagram is constructed using global thermodynamic equilibrium argument applied to the charged string and 5D electrically charged black hole. The results derived illustrate that electric charge makes the black string less stable.            
\end{abstract}

\pacs{04.20.Dw, 04.20.Cv, 04.70.Bw \hfill  
Alberta-Thy-03-12}
\maketitle

\section{\label{sec:level1}INTRODUCTION}

Spacetime with large compact extra dimensions admits a variety of
black objects. Study of topological phase transitions between these objects is a
subject of high interest. A well-known example is a uniform black
string in spacetime with compact extra dimension. The topology of
the horizon of the black string is different from the spherical
topology of a black hole solution, which is another possible phase. A
vacuum black string solution can be obtained from a lower dimensional,
spherically symmetric, static, vacuum black hole solution by adding flat compact extra dimension. 
In the case of more than one, say $p$, extra dimensions the solution is called the black p-brane. Simple analysis shows that for a given size of extra dimensions there exists a critical value of
mass $M_{cr}$. If $M>M_{cr}$, the entropy of the black string or black brane is
larger than the entropy of the black hole, while for $M<M_{cr}$ the
situation is opposite. 

Based on this simple argument, Gregory and Laflamme (GL) \cite{GL}
came to a conclusion that a black string (brane) must be unstable if its 
mass is smaller than some critical value. 
Studying S-wave gravitational perturbations of black strings they
discovered classical instability of these objects. Namely, they
obtained a dispersion relation between imaginary part $\Omega$ of the frequency,
and the wave number $k$ of the perturbation field mode, which
clearly illustrates the instability. The dispersion relation allows
one to find a rate of decay corresponding to a given wave number $k$. 
The instability starts at the special value $k_{cr}$. $\Omega$ vanishes for this threshold mode. Positive
values of $\Omega$ correspond to the smaller than $k_{cr}$ values of
the wave numbers. Finally, the zero-value wave number corresponds to
$\Omega=0$, that indicates stability of $D$-dimensional
Schwarzschild-Tangherlini black hole. 

Existence of the critical (maximal) wave number, $k_{cr}$, in the
instability spectrum corresponds to the minimal wave-length
$L_{cr}=2\pi/k_{cr}$ of the perturbation threshold mode. The modes
with smaller values of the wave-length are stable. This behavior is similar to the classical
Jeans instability. If there exist compact extra spatial dimensions with a size smaller than the minimal
wave-length, $L_{cr}$, then unstable modes can not fit into the compact dimensions, and the instability does
not arise. In the opposite case, a compactified black string (brane) solutions are unstable.

The GL instability is a generic property of black objects in spacetime with compact extra dimensions. This effect was studied by many authors. For example, in \cite{GL2} GL instability for a dilaton black string with magnetic charge was demonstrated. Instability in magnetically charged black string was studied in \cite{Miya}. It was found that for a certain range of magnetic charge values the string becomes stable, if compared with a neutral one. Instability in boosted strings was studied in \cite{Myers}. It was found that in the frame comoving with the string results are largely unchanged, if compared with a static black string. However, in the frame where the string is moving along its length, the instability strongly depends on the velocity of the string. For example, the threshold mode appears as a wave traveling with the boost velocity along the string. 

Numerical evolution of a perturbed, unstable black string was studied in \cite{Chop}. It was found that at an 
intermediate time of the evolution a non-uniform black string forms. This string reminds a distorted spherical black 
hole connected to a thin black string. Topological transition between black strings in 5D and 6D, and the corresponding Kaluza-Klein black holes was studied in \cite{Jutta}. It was found that the black string and the black hole phases merge at a topology changing transition. Large D asymptotics of a marginally stable black string were studied in \cite {KolS}. Detailed reviews on the phase transitions between black strings and black holes, and the GL instability issue are given in \cite{Kol} and \cite{HaNOb}, respectively.

There is an interesting relation between dynamical instability of a black brane and off-shell instability of the corresponding Euclidean black hole. For example, the threshold unstable mode of a neutral black brane corresponds to the Euclidean Schwarzschild black hole negative mode \cite{GPY}. Namely, $k^2_{cr}=-\lambda$, where $\lambda$ is the negative eigenvalue of the spectral problem related to the Euclidean black hole off-shell perturbations. General analysis of such relations is given in \cite{Reall}. Negative modes of 4D Reissner-Nordstr\"om black hole with magnetic charge corresponding to threshold unstable mode of 5D black string were found in \cite{MonSan}.
 
In a general case, the size of extra compact dimensions is spacetime parameter which
is fixed at asymptotically flat region. However, the proper length of
the extra compact dimensions at the vicinity of a black object
generally depends on the present matter and fields. The matter or
fields may increase or decrease the local size of the compact
dimensions changing the black object instability spectrum. This
effect is illustrated on several types of black strings.  For example,
presence of magnetic and dilaton fields tends to stabilize the black
string \cite{GL2}. Similar behavior of the critical wave-number for black string
with magnetic charge was observed in \cite{Miya}.                 

In this paper we shall study instability of electrically charged
5D black string. The corresponding dimensionally reduced solution is S-dual to
the dimensionally reduced solution for 5D black string with magnetic charge \cite{Miya}. 
We consider static S-wave perturbation of
the charged black string and search for the threshold mode. Existence of the
threshold mode implies existence of the instability spectrum. We construct a critical curve in topological phase transition diagram. The curve corresponds to a non-uniform black string and separates the charged black string phase and the corresponding charged 
Kaluza-Klein black hole phase. The shape of the curve illustrates that electric charge
tends to destabilize the black string. Similar behavior can be
inferred from the global thermodynamic equilibrium condition between the charged black string and charged black hole.

The paper is organized as follows. In Section II we present
the metric of the 5D electric black string and discuss its
properties. In Section III we consider static S-wave perturbation of
the charged string and derive the corresponding master equation. In
Section IV we integrate numerically the master equation and construct
the critical curve in the topological phase transition diagram. In
Section V we derive a similar curve using the global thermodynamic
equilibrium argument. In Section VI we discuss and compare the phase diagrams for the electric and magnetic black 
strings. Section VII contains discussion of our results. In this paper we shall use the
following convention of units $16\pi G_{(4)}=c=\hbar=k_B=1$. Spacetime signature is $+(D-2)$.

\section{5D charged black string solution}

\subsection{5D theory}

In this Section we describe a solution for 5D electrically charged, compactified
black string.  Let us consider the following action 
\be\n{10}
S=\frac{1}{16\pi G_{(5)}}\int_0^L dz\int dx^4\sqrt{-g}\left(R-\frac{1}{4}F^2\right),
\ee
where 
\be\n{10a}
F_{\mu\nu}=\nabla_{\mu}A_{\nu}-\nabla_{\nu}A_{\mu}.
\ee
Here $A_{\mu}$ is the 5D electromagnetic vector potential, and $L$ is the size of compact dimension. 
The corresponding Einstein-Maxwell equations read 
\ba\n{11}
&&R_{\mu\nu}-\frac{1}{2}g_{\mu\nu}R=\frac{1}{2}T^{(em)}_{\mu\nu},\\
&&\nabla_{\nu}F^{\mu\nu}=0\hhh \nabla_{[\lambda}F_{\mu\nu]}=0.
\ea
The 5D electromagnetic field energy-momentum tensor is given by
\ba\n{12}
T^{(em)}_{\mu\nu}&=&F_{\mu}^{\,\,\,\lambda}F_{\nu\lambda}-\frac{1}{4}g_{\mu\nu}F^2.
\ea
Here and in what follows $\nabla_{\mu}$ denotes a covariant derivative defined with respect to 5D metric, whereas 
coma stands for a partial derivative. 

In a general case, electric charge $Q$ associated with d-form $F_{(d)}$ is defined as follows
\ba\n{31}
Q&=&(-1)^{1+d(D-d)}\int_{V_{(D-d)}}\star F_{(D-d)},
\ea
where the Hodge dual to $F_{(d)}$ is defined by 
\ba\n{9}
\star F^{\mu_1\ldots\mu_{(D-d)}}=\frac{\varepsilon^{\mu_1\ldots\mu_{(D-d)}\nu_1\ldots\nu_{d}}}{d!\sqrt{-g}}F_{\nu_1
\ldots\nu_d}.
\ea
Magnetic charge $P$ is defined by
\ba\n{9a}
P&=&\int_{V_{(d)}}F_{(d)}.
\ea
The integral over $d$-form $F_{(d)}$ and the corresponding volume element $V_{(d)}$ are defined as follows
\ba\n{32}
\int_{V_{(d)}}F_{(d)}&=&\int_{V_{(d)}}F_{|\mu_1...\mu_d|}dx^{\mu_1}...dx^{\mu_d}\hhh V_{(d)}=\int_{V_{(d)}}\star1,\nonumber\\
\ea
where $|\mu_1...\mu_d|=\mu_1<...<\mu_d$ means proper orientation, and $\varepsilon^{|\mu_1\ldots\mu_{(D-d)}\nu_1\ldots\nu_{d}|}=+1$. The definition of electric charge corresponds to the Heaviside-Lorentz system of units. In our case $d=2$.

The Komar mass of a black object is defined by
\be\n{M}
M=\frac{1}{16\pi G_{(D)}}\frac{D-2}{D-3}\int_{V_{(D-2)}^{\infty}}d^{D-2}\Sigma_{\mu\nu}\nabla^{\mu}k^{\nu},
\ee
where $k^{\mu}$ is the timelike Killing vector normalized at infinity as $k^{\mu}k_{\mu}=-1$. Another definition of mass of a black object in spacetime with compact dimensions was given in \cite{HaOb}.

For spacetime with compact extra dimensions of size $L$ the $D$-dimensional and $4D$ gravitational constants are related as follows
\be\n{G}
G_{(D)}=G_{(4)}L^{D-4}.
\ee
 
We consider 5D metric of the following form
\be\n{16}
ds^2=-f_1dt^2+\frac{dr^2}{f_1f_2}+f_2dz^2+f_3d\Omega^2_{(2)},
\ee
where $f_i=f_i(r)$, $i=1,2,3$, $d\Omega^2_{(2)}=d\theta^2+\sin^2\theta d\phi^2$, is the metric on a unit 2D round sphere, and $z$ is the coordinate of the compact dimension of size $L$.
In the next subsection we shall see that this metric can be derived by oxidizing (uplifting) a 4D dimensional solution, 
which does not depend on compact coordinate $z$, to 5D spacetime of the charged black string \cite{Ortin}. Such 4D 
solution is related to solutions of the so-called {\em a} -model. In our case this is 4D electrically charged 
dilaton black hole.

\subsection{The {\em a} -model}

Here we discuss the 4D {\em a} -model solution representing dilaton black hole with electric charge. Let us start 
from the 4D {\em a} -model action \cite{Ortin}
\be\n{1}
\bar{S}=\frac{1}{16\pi G_{(4)}}
\int d^4x\sqrt{-\bar{g}}\left(\bar{R}-\frac{1}{2}(\bar{\nabla}\varphi)^2
-\frac{1}{4}e^{-2a\varphi}\bar{F}^2\right), 
\ee
where $\bar{R}$ is the Ricci scalar of the $4D$ spacetime, $\varphi$ is the dilaton field, and $a$ is the dilaton-electromagnetic field coupling constant. The electromagnetic field tensor is
\be\n{1a} 
\bar{F}_{ij}=\bar{\nabla}_{i}\bar{A}_{j}-\bar{\nabla}_{j}\bar{A}_{i},
\ee
where $\bar{A}_{i}$ is the 4D electromagnetic vector potential. Here and in what follows $\bar{\nabla}_i$ denotes a
covariant derivative with respect to the $4D$ metric, and $i,j,k=0,1,2,3$.

The Einstein-Maxwell-dialton equations read
\ba\n{2}
&&\bar{R}_{ij}-\frac{1}{2}\bar{g}_{ij}\bar{R}=\frac{1}{2}\left(\bar{T}^{(d)}_{ij}+e^{-2a\varphi}\bar{T}^{(em)}_{ij}\right),\\
&&\bar{\nabla}_{j}\left(e^{-2a\varphi}\bar{F}^{ij}\right)=0\hhh \bar{\nabla}_{[k}\bar{F}_{ij]}=0,\\
&&\bar{\nabla}^2\varphi+\frac{a}{2}e^{-2a\varphi}\bar{F}^2=0.
\ea
Here the dialton and the 4D electromagnetic field energy-momentum tensors are given by 
\ba\n{3}
\bar{T}^{(d)}_{ij}&=&(\bar{\nabla}_{i}\varphi)(\bar{\nabla}_{j}\varphi)-\frac{1}{2}\bar{g}_{ij}(\bar{\nabla}\varphi)^2,\\
\bar{T}^{(em)}_{ij}&=&\bar{F}_{i}^{\,\,\,k}\bar{F}_{jk}-\frac{1}{4}\bar{g}_{ij}\bar{F}^2,
\ea
respectively.

We are interested in static, spherically symmetric solutions of the equations. One of such solutions is a dilaton black 
hole with electric charge. The corresponding metric is \cite{Ortin}
\be\n{4}
d\bar{s}^2=-WH^{-v}dt^2+W^{-1}H^{v}dr^2+r^2H^{v}d\Omega^2_{(2)},
\ee
where $v=2/(1+4a^2)$, and
\be\n{5}
W=1-\frac{w}{r}\hhh H=1+\frac{h}{r}\hhh w=\frac{h}{4}[b^2(1+4a^2)-1].
\ee
The electromagnetic vector potential and the dilaton field are given by
\ba\n{6}
\bar{A}_{i}&=&b(H^{-1}-1)\delta_{i}^{\,\,\,t}\hhh \varphi=-2av\ln H,
\ea
respectively. In a particular case of zero coupling constant, $a=0$, the solution is the 4D Reissner-Nordstr\"om black hole. 

To proceed with the oxidation \cite{Ortin} let us consider 5D metric \eq{16} in the following form
\be\n{13}
ds^2=e^{-\tfrac{\varphi}{\sqrt{3}}}d\bar{s}^2+e^{\tfrac{2\varphi}{\sqrt{3}}}dz^2,
\ee
where $d\bar{s}^2$ is 4D metric \eq{4}.
The following relations between the 5D and 4D objects hold
\ba\n{14}
R&=&e^{\tfrac{\varphi}{\sqrt{3}}}\left(\bar{R}-\frac{1}{2}(\bar{\nabla}\varphi)^2+\sqrt{3}\bar{\nabla}^2\varphi\right),\\
\sqrt{-g}&=&e^{-\tfrac{\varphi}{\sqrt{3}}}\sqrt{-\bar{g}}\hhh F^2=e^{\tfrac{2\varphi}{\sqrt{3}}}\bar{F}^2.
\ea
Applying these relations to action \eq{10}, integrating over $z$, and using the relationship \eq{G} between the gravitational constants in 5D and 4D we derive
\be\n{15}
S=\frac{1}{16\pi G_{(4)}}\int dx^4\sqrt{-\bar{g}}\left(\bar{R}-\frac{1}{2}(\bar{\nabla}\varphi)^2-\frac{1}{4}e^{\tfrac{\varphi}{\sqrt
{3}}}\bar{F}^2\right).
\ee 
This is exactly 4D action \eq{1} with $a=-1/(2\sqrt{3})$. 

\subsection{5D electrically charged black string}

Oxidation of the dilaton black hole with electric charge to 5D gives us the electric black string solution \eq{16} with
\ba\n{17}
f_1&=&WH^{-2}\hhh f_2=H\hhh f_3=r^2H,\\
A_{\mu}&=&b(H^{-1}-1)\delta_{\mu}^{\,\,\,t}\hhh w=\frac{h}{3}(b^2-3),\n{17a}
\ea
where $W$ and $H$ are given in \eq{5}.
This solution coincides with that of \cite{Jutta}, p. 19, for $a=0$, $h=r_0\,\sinh^2\beta$, $w=r_0$, $b=\sqrt{3}\,\coth
\beta$, and $A_{\mu}\rightarrow A_{\mu}+\sqrt{3}\,\tanh\beta\,\delta_{\mu}^{\,\,\,t}$.

Note that for the 5D black string and the 4D black hole solutions $r>0$, $h>0$, and the corresponding event horizons are defined by $r=w>0$. The parameters $h$ and $w$ can be expressed through mass 
and electric charge of the black string. 

The definition \eq{M} and relation \eq{G} give us the black string mass, which in our units is
\ba\n{18}
M&=&6\pi(w+2h).
\ea
Using the definition of electric charge \eq{31} for $d=2$, we derive
\ba\n{33}
Q&=&4\pi Lbh.
\ea
Expressions \eq{18} and \eq{33} give
\ba\n{34}
w&=&\frac{1}{6\pi}\sqrt{M^2-3(Q/L)^2},\\
h&=&\frac{1}{12\pi}\left(M-\sqrt{M^2-3(Q/L)^2}\right).
\ea
Thus, the magnitude of the black string electric charge is defined within the range
\ba\n{35}
0\leqslant|Q|<\frac{ML}{\sqrt{3}}.
\ea

\section{Static perturbations of the black string} 

\subsection{S-wave static perturbations}

To study GL instability of 5D charged black strings we consider gravitational perturbations $h_{\mu\nu}\ll1$ of the metric $g_{\mu\nu}$, given in \eq{16}, \eq{17}. In general, for the 5D metric we have 15 components of the perturbation field $h_{\mu\nu}
$. Following \cite{GL} we consider S-wave perturbations only. In this case the perturbation field has 7 components
\be\n{40}
h_{\mu\nu}=\{h_{tt},h_{tr},h_{tz},h_{rr},h_{rz},h_{\theta\theta},h_{\phi\phi}=h_{\theta\theta}\sin^2\theta,h_{zz}\}.
\ee
Such gravitational perturbation in general induces perturbation in the electromagnetic field $F_{\mu\nu}$. In the case of the electric black string, $F_{\mu\nu}$ has in general the following induced components 
\be\n{40a}
\delta F_{\mu\nu}=\{\delta F_{tr},\delta F_{tz},\delta F_{rz}\},
\ee
which depend on $(t,r,z)$ coordinates.

Fourier mode of the S-wave propagating in $z$ direction along the black string reads
\be\n{41}
h_{\mu\nu}=\Re\left\{a_{\mu\nu}(r)e^{-i\omega t+ikz}\right\}.
\ee
This mode is unstable if $\omega=i\Omega$, where $\Omega>0$. Threshold of GL instability corresponds to time-independent (critical) mode with $\Omega=0$. We shall study static S-wave perturbations of the black string and 
search for the critical GL mode.  

In the case of static S-wave as a result of the symmetry $t\rightarrow -t$, the number of the metric perturbation field 
components reduces to 5
\be\n{41a}
h_{\mu\nu}=\{h_{tt},h_{rr},h_{rz},h_{\theta\theta},h_{\phi\phi}=h_{\theta\theta}\sin^2\theta,h_{zz}\}.
\ee
The perturbed electromagnetic field \eq{40a} has $\delta F_{tr}$ and $\delta F_{tz}$ components only, which depend on $(r,z)$ coordinates. Using the electromagnetic gauge freedom the corresponding vector potential can be cast into the form 
$A_{\mu}=A_t(r,z)\delta_{\mu}^{\,\,\,t}$.

Thus, we can present perturbed 5D charged black string solution \eq{16}, \eq{17}, \eq{17a} as follows
\be\n{42}
ds^2=-f_1e^{2\tau}dt^2+\frac{e^{2\sigma}dr^2}{f_1f_2}+f_2e^{2\beta}(dz-\alpha dr)^2+f_3e^{2\gamma}d\Omega^2_{(2)},
\ee
\be\n{42a}
A_{\mu}=[A_t(r)+a_t(r,z)]\delta_{\mu}^{\,\,\,t}.
\ee
Here the components $\{\tau,\sigma,\beta,\alpha,\gamma,a_t\}\ll1$ are functions of $r$ and $z$ only.
     
Because of the freedom in coordinate choice in general relativity, there is a gauge freedom in the perturbation field. To fix the gauges we have to fix infinitesimal diffeomorphisms $\xi^{\mu}$ in the coordinate transformations $x^{\mu}=x^{\mu'}+\xi^{\mu}$. In our case $\xi^{\mu}=[\xi^{r}(r,z),\xi^{z}(r,z)]$. These diffeomorphisms induce gauge transformations in the metric $g_{\mu\nu}$, and in the vector potential $A_{\mu}$ as follows
\ba\n{43}
\delta g_{\mu\nu}&=&\left(\mathcal{L}_{{\bf \xi}}{\bf g}\right)_{\mu\nu}=\nabla_{\mu}\xi_{\nu}+\nabla_{\nu}\xi_{\mu},\\
\delta A_{\mu}&=&\left(\mathcal{L}_{{\bf \xi}}{\bf A}\right)_{\mu}=A_{\mu,\nu}\xi^{\nu}+A_{\nu}\xi^{\nu}_{\,\,,\mu}.
\ea
As a result, the components of the perturbation field transform in the following way
\ba\n{44}
\delta\tau&=&\frac{1}{2}\xi_rf_{1,r}f_2\hhh \delta\sigma=\xi_{r,r}f_1f_2+\frac{1}{2}\xi_r (f_1f_2)_{,r},\\
\delta\gamma&=&\frac{1}{2}\xi_rf_1f_2f_3^{-1}f_{3,r}\hhh \delta\beta=\xi_{z,z}f_2^{-1}+\frac{1}{2}\xi_rf_1f_{2,r},\,\,\,\,\,\,
\,\,\,\n{44a}\\  
\delta\alpha&=&f_2^{-2}(\xi_zf_{2,r}-\xi_{z,r}f_2-\xi_{r,z}f_2).\n{44b}
\ea
In the case of the 5D electric black string the only non-vanishing induced electromagnetic perturbation, $a_t$, transforms as   
\ba\n{45}
\delta a_t=\xi_rf_1f_2A_{t,r}.
\ea
$a_t$ corresponds to the perturbation of timelike (electrostatic) component of $A_{\mu}$. 

Our system of equations contains 6 `field' variables
$\{\tau,\sigma,\beta,\alpha,\gamma,a_t\}$ with the gauge freedom
\eq{44}-\eq{45} generated by 2 functions $\xi^r$ and $\xi^z$ of 2
variables. As usual for such a case, there exist $6-2\cdot2=2$
`physical' degrees of freedom (see e.g. \cite{Dirac}). We shall demonstrate 
that one can choose $\gamma$ and $a_t$ as such `physical' degrees
of freedom. Namely, we show that these objects obey 2 decoupled,
second order ordinary differential equations. As we show later, one
of the equations (for $a_t$) does not have unstable modes, while the other
one, the master equation for $\gamma$, is responsible for existence of the GL unstable
threshold mode.

\subsection{Master equation for the electric black string}

We use the same gauge as in \cite{Miya}. Namely, we define
$\tau=\tau'+\delta\tau=0$, that fixes $\xi^r$ through equation
\eq{44}. This gauge preserves the gravitational redshift function $f_1$ under
the perturbation field. The next gauge choice is
$\beta=\beta'+\delta\beta=0$, that fixes $\xi^{z}$ through equation
\eq{44a}. This gauge preserves the metric along $z$ coordinate. Thus, 
we are left with $\{\sigma,\alpha,\gamma,a_t\}$. Assuming that the amplitudes in the
Fourier modes \eq{41} of the perturbation field are real valued
functions, we have
\ba\n{46}
\sigma&=&\tilde{\sigma}(r)\cos(kz)\hhh \alpha=-\tilde{\alpha}(r)k\sin(kz),\\
\gamma&=&\tilde{\gamma}(r)\cos(kz)\hhh a_t=\tilde{a}_t(r)\cos(kz).
\ea
Here the form of the Fourier mode of $\alpha$ reflects the fact that $\delta\alpha$ is an antisymmetric function of $z$.

Our goal is to derive the master equation corresponding to the perturbation field. Currently we have three components of the gravitational perturbation field, $\{\sigma,\alpha,\gamma\}$, and one component of the induced electrostatic perturbation, $a_t$. They solve system of the Einstein-Maxwell equations.          
      
We employed GRtensorII package to derive a reduced system of the Einstein-Maxwell equations for the first order 
gravitational perturbation of metric \eq{16}, \eq{17}, \eq{17a} corresponding to the electric black string. As a result, we have five components, $\{tt,rr,rz,\theta\theta,zz\}$, of the Einstein equations and one, $\{t\}$, component of the Maxwell equations, $\nabla_{\nu}F^{\mu\nu}=0$. The Maxwell equations $\nabla_{[\lambda}F_{\mu\nu]}=0$ are satisfied identically. Using $\{rz\}$ component of the Einstein equations we can express $\tilde{\sigma}$ in terms of $\tilde{\gamma}$ and $\tilde{a}_t$, 
\be
\tilde{\sigma}=\frac{4(r-w)(r\tilde{\gamma}_{,r}+\tilde{\gamma})-bh\tilde{a}_t}{4r-3w},
\ee
substitute the result into $\{rr\}$ component of the Einstein equations, and solve it for $\tilde{\alpha}$ in terms of $\tilde{\gamma}$ and $\tilde{a}_t$,
\ba
\tilde{\alpha}&=&-\frac{2w(2r-3w)\tilde{\gamma}_{,r}-[4w-4k^2r^2(4r-3w)]\tilde{\gamma}}{k^2(4r-3w)^2}\nonumber\\
&+&\frac{bh[(4r-3w)\tilde{a}_{t,r}+4\tilde{a}_t]}{k^2(4r-3w)^2}.
\ea
Thus, we can express $\tilde{\sigma}$ and $\tilde{\alpha}$ in terms of $\tilde{\gamma}$ and $\tilde{a}_t$. Substituting the result into remaining Einstein-Maxwell equations we see that $\{zz\}$ component is satisfied identically, whereas the components $\{tt\}$ and $\{\theta\theta\}$ together with $\{t\}$ component of the Maxwell equation represent a system of three mutually compatible equations for $\tilde{\gamma}$ and $\tilde{a}_t$. Analysis of the equations shows that we can eliminate $\tilde{\gamma}$ and derive a single equation for $\tilde{a}_t$
\ba\n{47}
\frac{r-w}{r}\tilde{a}_{t,rr}+\frac{2(r-w)(w+2h)}{r(wr+2hr-wh)}\tilde{a}_{t,r}&=&k^2\tilde{a}_t.
\ea
Thus, the electrostatic perturbation decouples from the gravitational one. As we illustrate in Appendix, there are no 
unstable threshold modes. Thus we can take $\tilde{a}_t=0$. 

As a result, $\{tt\}$ and $\{\theta\theta\}$ components of the Einstein equations and $\{t\}$ component of the Maxwell 
equations become equivalent and give us the master equation for the static S-wave gravitational perturbation of 
the electric black string
\ba\n{50}
\frac{r-w}{r}\tilde{\gamma}_{,rr}-\frac{w(2r-3w)}{r^2(4r-3w)}\tilde{\gamma}_{,r}+\frac{2w}{r^2(4r-3w)}\tilde{\gamma}
&=&k^2\tilde{\gamma}.\nonumber\\
\ea
Remarkably, this equation has only one parameter $w$, whereas the black string metric \eq{16}, \eq{17}, \eq{17a} has two, $w$ and $h$.

\section{Numerical analysis}

We shall integrate the master equation \eq{50} numerically, applying the shooting method \cite{Fort}. To proceed we 
map the semi-infinite interval $r\in[w,+\infty)$ into finite one and rewrite the master equation in dimensionless form 
using the transformations
\ba\n{51}
r&=&\frac{w}{1-x}\hhh k=\frac{\kappa}{w},
\ea 
where $x\in[0,1]$. Here $x=0$ corresponds to the black string event horizon. The master equation takes the form
\ba\n{52}
x^2\tilde{\gamma}_{,xx}+xP(x)\tilde{\gamma}_{,x}+Q(x)\tilde{\gamma}&=&0,
\ea
where
\ba\n{53}
P(x)&=&-\frac{(3x^2+6x-1)}{(1-x)(1+3x)},\\
Q(x)&=&\frac{2x}{(1-x)(1+3x)}-\frac{x\kappa^2}{(1-x)^4}.
\ea
This equation has a regular singular point at $x=0$, and irregular singular point at $x=1$. We are looking for regular 
solutions which are finite everywhere together with their derivatives and vanish at infinity. Namely, $\tilde{\gamma}(1)=0$. Applying the method of Frobenius we can construct approximate solution near the horizon. The solution reads
\ba\n{54}
\tilde{\gamma}(x)\approx c_0-c_0(2-\kappa^2)x+\cdots,
\ea 
where $c_0$ is arbitrary constant, which we take equal to one. From \eq{54} we derive the following boundary conditions on the black string event horizon
\ba\n{55}
\tilde{\gamma}(0)=1\hhh \tilde{\gamma}_{,x}(0)=\kappa^2-2.  
\ea
Using the boundary conditions and implementing the shooting code written in FORTRAN we found $\kappa_{cr}
\approx 0.876$, for the critical GL mode. This result coincides with that of \cite{Myers} for the neutral 5D black string. 

To obtain a critical wave number, $k_{cr}$, we use transformations \eq{51}
\ba\n{66}
k_{cr}&=&\frac{\kappa_{cr}}{w(M,Q)}.
\ea
Here $w(M,Q)$ is given by \eq{34}. The critical unstable mode corresponds to the highest mode in the instability 
spectrum $\Omega=\Omega(k)$, i.e. to the lowest frequency $\Omega=0$. For the second order equations for static 
spacetime \eq{16}, \eq{17}, \eq{17a} the dispersion relation $\Omega=\Omega(k)$ has two roots $k=0$ and $k=k_{cr}$. The region $k\in[0,k_{cr}]$ defines the instability spectrum. If we set the size of the compact dimension $L=L_{cr}\equiv2\pi/k_{cr}$, then for such spacetime (with fixed $M$ and $Q$)  there will be only one unstable mode, the critical one. For spacetime with $L>L_{cr}$ additional unstable modes are possible, and for spacetime with $L<L_{cr}$ there are no unstable modes at all. Relation \eq{66} can be presented in the following dimensionless form
\ba\n{67}
\mu&=&\sqrt{9\kappa_{cr}^2+3q^2}.
\ea
Here we introduced dimensionless parameters of mass $\mu=M/L$ and electric charge $q=Q/L^2$. 

\begin{figure}[htb]
\begin{center} 
\includegraphics[height=6.03cm,width=6cm]{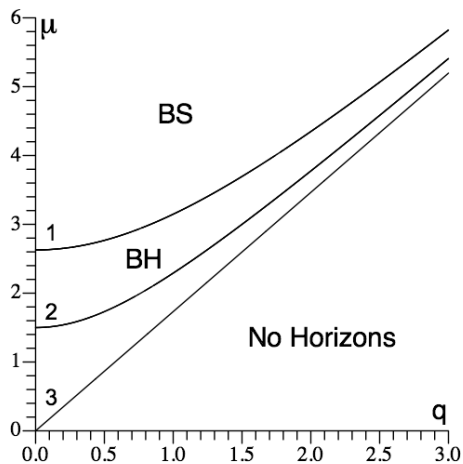} 
\caption{The dynamical critical curve (1) for the electric black string (BS) $-$ electric black hole (BH) topological phase transition 
corresponding to $\kappa_{cr}\approx\ 0.876$. The thermodynamical curve (2) corresponds to \eq{77}. Line (3), $\mu=q\sqrt{3}$, corresponds to the extreme value of the electric charge \eq{35}, \eq{75}. Here $\mu=M/L$, $q=Q/L^2$. Curves (1) and (2) approach asymptotically line (3).} \label{f1} 
\end{center}
\end{figure}

Line (1) in Figure 1 represents this relation for dimensionless critical wave number $\kappa_{cr}\approx 0.876$. We 
see that addition of electric charge to a neutral 5D black string makes it less stable. Namely, for a given mass $M>M_
{cr}$ corresponding to a stable neutral black string, gradual addition of electric charge  shifts the black string close to 
the critical curve, and for $Q>Q_{cr}(M_{cr})$ the black string becomes unstable.

\section{Thermodynamics of the black strings}

Black strings are higher dimensional objects which have event horizon. Thus, as in the case of a black hole, they can be considered as thermodynamical systems which have entropy and temperature. The entropy is defined as follows
\ba\n{68}
S&=&\frac{A_H}{4G_{(D)}},
\ea
where $A_H$ is the event horizon surface area. For the electric black string we have
\ba\n{69}
S_{EBS}&=&16\pi^2\sqrt{w}(w+h)^{\tfrac{3}{2}}=\frac{\sqrt{2}}{9}\left(M^2-3(Q/L)^2\right)^{\tfrac{1}{4}}\nonumber\\
&\times&\left(M+\sqrt{M^2-3(Q/L)^2}\right)^{\tfrac{3}{2}}.
\ea
We assume that unstable 5D electric black string may undergo a topological phase transition into electrically charged 5D black hole. We can compare entropy \eq{69} with that of the electrically charged black hole $S_{EBH}$, and define micro-canonical equilibrium condition for the electric black string as $S_{EBS}>S_{EBH}$. The corresponding critical curve is defined as $S_{EBS}=S_{EBH}$. 

Unfortunately, an exact solution of the 5D electrically charged black hole in spacetime with one compact dimension is 
unknown. However, we can make an estimate for the micro-canonical critical curve by comparing the electric black string entropy, $S_{EBS}$, with that of the 5D Reissner-Nordstr\"om black hole. We put mass and electric charge of the black hole equal to those of the black string and use relation \eq{G}. 

The 5D Reissner-Nordstr\"om black hole metric is \cite{Ortin}
\ba\n{70}
ds^2&=&-\tilde{W}\tilde{H}^{-2}dt^2+\tilde{W}^{-1}\tilde{H}dr^2+r^2\tilde{H}d\Omega_{(3)}^2,\\
\tilde{W}&=&1-\frac{\tilde{w}}{r^2}\hhh \tilde{H}=1+\frac{\tilde{h}}{r^2}\hhh \tilde{w}=\frac{\tilde{h}}{3}(\tilde{b}^2-3),
\ea
where $d\Omega_{(3)}^2$ is the metric on a unit 3D round sphere, and the electromagnetic vector potential is given by
\ba\n{72}
A_{\mu}&=&\tilde{b}(\tilde{H}^{-1}-1)\delta_{\mu}^{\,\,\,t}.
\ea
The black hole horizon is located at $r=\sqrt{\tilde{w}}$. The Komar mass \eq{M} and electric charge \eq{31} are
\ba\n{73}
M&=&\frac{6\pi^2}{L}(\tilde{w}+2\tilde{h})\hhh Q=4\pi^2\tilde{h}\tilde{b}.
\ea
Thus, we have
\ba\n{74}
\tilde{w}&=&\frac{L}{6\pi^2}\sqrt{M^2-3(Q/L)^2},\\
\tilde{h}&=&\frac{L}{12\pi^2}\left(M-\sqrt{M^2-3(Q/L)^2}\right).
\ea
Hence, the black hole electric charge is defined within the range
\ba\n{75}
0\leqslant|Q|<\frac{ML}{\sqrt{3}}.
\ea
For the 5D Reissner-Nordstr\"om black hole we have
\ba\n{76}
S&=&\frac{8\pi^3}{L}(\tilde{w}+\tilde{h})^{\tfrac{3}{2}}=\frac{L^{\tfrac{1}{2}}}{3\sqrt{3}}\left(M+\sqrt{M^2-3(Q/L)^2}\right)^{\tfrac{3}{2}}.\nonumber\\
\ea
According to expressions \eq{35} and \eq{75} electric charge of the black string and the black hole is defined in the same range. Approximating $S_{EBH}$ with expression \eq{76} we derive the micro-canonical critical curve. In dimensionless form the equation for the curve is
\ba\n{77}
\mu&=&\sqrt{9/4+3q^2}.
\ea
This curve is shown on Figure 1 (curve (2)) together with the critical curve \eq{67} of the dynamical perturbation. These curves are hyperbolae. They have qualitatively similar behavior and illustrate that electrically charged black string is less stable than the neutral one. Let us note that line (1) for the dynamical instability is always higher than line (2).

\section{5D magnetic black string}

It is interesting to compare the obtained phase diagram (Figure 1) with a similar diagram for 5D magnetically charged 
black string. Let us remind some properties of the 5D magnetic black string, which was studied (in the case of general D) in \cite{Miya}. The magnetic black string metric can be derived as follows. Dimensionally reduced magnetic black 
string gives 4D dilaton black hole with magnetic charge which is another solution to the 4D {\em a} -model discussed 
above. This solution is S-dual to the 4D dilaton black hole with electric charge, \eq{4}-\eq{6}.  Applying S-duality transformation \cite{Ortin}
\ba\n{78}
F'&=&e^{-2a\varphi}\star F\hhh \varphi'=-\varphi,
\ea
to the 4D dilaton black hole with electric charge one derives the electromagnetic potential and the dilaton field
\ba\n{79}
A_{\mu}&=&bh\cos\theta\delta_{\mu}^{\,\,\,\phi}\hhh \varphi=2av\ln H,
\ea
of the 4D dilaton black hole with magnetic charge. The duality transformation does not change 4D metric \eq{4}. Oxidizing the metric to 5D we derive the magnetic black string solution \eq{16} with 
\ba\n{80}
f_1&=&WH^{-1}\hhh f_2=H^{-1}\hhh f_3=r^2H^2,\\
A_{\mu}&=&bh\cos\theta\delta_{\mu}^{\,\,\,\phi}\hhh w=\frac{h}{3}(b^2-3),
\ea
where $W$ and $H$ are given in \eq{5}. This solution coincides with that of \cite{Miya}, p. 381, for $h=r_-$, $w=r_+-r_-$, $r\rightarrow r-r_-$, and $b=\sqrt{3r_+/r_-}$. Using \eq{M}, \eq{G} and \eq{9a} we derive mass $M$ and magnetic charge $P$ of the black string
\ba\n{81}
M&=&6\pi(w+h)\hhh P=-4\pi bh.
\ea

\begin{figure}[htb]
\begin{center} 
\includegraphics[height=4.83cm,width=7.5cm]{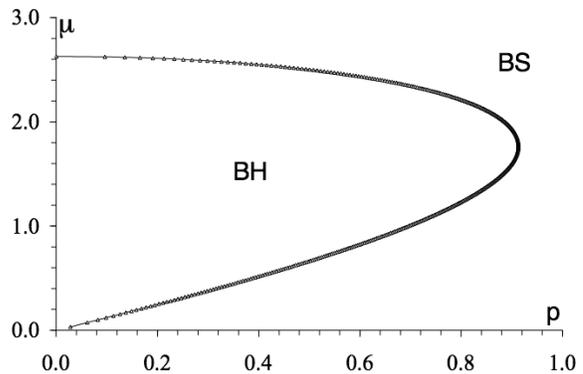} 
\caption{The dynamical critical curve for the magnetic black string (BS) $-$ black hole (BH) topological phase transition. Here $\mu=M/L$, $p=P/L$.} \label{f2} 
\end{center}
\end{figure}

We put the magnetic black string mass equal to the mass of the electric black string. Using expressions \eq{81} we 
have
\ba\n{83}
w&=&\frac{1}{6\pi M}\left(M^2-3/4P^2\right),\\
h&=&\frac{P^2}{8\pi M}.\n{83a}
\ea

To present the critical mode of the magnetic black string in $(M,P)$ parameter space we have to construct a relation between $\mu=M/L$ and $p=P/L$ for $L=L_{cr}$. We use \eq{83}, \eq{83a} and the numerical results of \cite{Miya}. The relation has the following parametric form
\be\n{92}
\mu=\frac{3\kappa_{cr}(m)}{1-m}\hhh p=\frac{2\kappa_{cr}(m)\sqrt{3m}}{1-m},
\ee
where 
\be\n{93}
m=\frac{h}{h+w}.
\ee
Figure 2 illustrates relation \eq{92}. We see that in contrast to electric charge, magnetic charge tends to stabilize black string.

We do not know the final state of the topological phase transition of the magnetically charged black string. A 5D magnetic black hole, dual to the 5D Reissener-Nordstr\"om black hole, does not belong to the same theory. Namely, the corresponding electromagnetic tensor is a 3-from, whereas for the 5D magnetic black string considered above it is a 2-form. One may propose a scenario where the final state of the topological phase transition is a neutral black hole pierced by magnetically charged string, which wraps around the compact dimension. Such a scenario tacitly assumes that topology of the magnetic charge 
is stable. 

\section{Discussions}

Thermodynamical and dynamical correspondence observed for the electric black string (see Figure 1) shows that as in the case of a neutral black string discussed in \cite{GL}, one may expect a decay of an electric black string into the 
corresponding Kaluza-Klein black hole. But as it is in the case of a neutral black string, the question whenever such 
transition is possible remains open. One may expect that because such transition is associated with extreme 
spacetime curvatures of a non-uniform black string horizon at the pinch regions, to describe such a transition quantum gravity considerations are necessary.

Our results show that, in contrast to the magnetic black string, the electric black string is less stable than a neutral one. Namely, electric charge tends to destabilize black string, whereas magnetic charge makes it more stable. This can be deduced from the form of $g_{zz}$ components of the corresponding metrics. For the electric black string $g_{zz}=H\geqslant1$, that makes the proper length along the compact dimension greater than $L_{cr}$. As a result, the electric black string is thinner than a neutral one, and hence less stable. For the magnetic black string $g_{zz}=H\leqslant1$, that makes the proper length along the compact dimension smaller than $L_{cr}$, and as a result, the string is thicker than a neutral one, and hence more stable.

S-duality \eq{78} between dimensionally reduced versions of the electric and magnetic black strings is broken after oxidization. In fact, we have such symmetry only in 4D for black holes, in 6D for black strings, etc. \cite{Ortin}.  It is interesting to study if there remains any connection between the instability spectrum properties of the electric and magnetic black strings induced by this S-duality. 

Finally, it would be interesting to analyze the relation between the threshold unstable mode of the electrically charged black string and negative mode of the corresponding Euclidean dilaton black hole with electric charge. To do this one can follow the procedure given in \cite{Reall}. However, our gauge is not suitable for such analysis.

\begin{acknowledgments}

One of the authors (V. F.) thanks the Natural Sciences and Engineering
Research Council of Canada and the Killam Trust for their support. We would like to thank Dan Gorbonos for bringing to our attention paper \cite{Miya}. 

\end{acknowledgments}

\appendix

\section{electrostatic perturbation}

Here we discuss existence of threshold modes in the electrostatic perturbation equation \eq{47}. Using 
transformations \eq{51} and \eq{93} we present the equation in the form
\be\n{A1}
x^2\tilde{a}_{t,xx}+\frac{2mx^2}{1+mx}\tilde{a}_{t,x}-\frac{\kappa^2x}{(1-x)^4}\tilde{a}_t=0.
\ee
Applying the method of Frobenius we derive the following boundary conditions
\ba\n{A2}
\tilde{a}_t(0)=0\hhh \tilde{a}_{t,x}(0)=\kappa^2\hhh \tilde{a}_t(1)=0.  
\ea
Using the boundary conditions and implementing the shooting code written in FORTRAN \cite{Fort} we found no indication of unstable threshold modes. 

This result can be inferred from analytical consideration of the
equation. Let us present \eq{A1} in the self-adjoint form
\be\n{A3}
\left((1+mx)^2\tilde{a}_{t,x}\right)_{,x}-\frac{\kappa^2(1+mx)^2}{x(1-x)^4}\tilde{a}_t=0.
\ee  
Multiplying this equation by $\tilde{a}_t$, integrating the result by parts, and using the boundary conditions \eq{A2} we 
derive
\be\n{A4}
\int_0^{1}\left((1+mx)^2\tilde{a}_{t,x}^2+\frac{\kappa^2(1+mx)^2}{x(1-x)^4}\tilde{a}_t^2\right)dx=0.
\ee
This equation has nontrivial solution $\tilde{a}_t$ only if $\kappa^2<0$. Thus, there are no unstable threshold modes.

\end{document}